\documentclass{iau}

\usepackage{amsmath}
\usepackage{graphicx}
\usepackage{multirow}

\begin{document}
\lefttitle{Di Cintio et al.}
\righttitle{IMRIs and relativistic dynamical friction}

\jnlPage{1}{4}
\jnlDoiYr{2025}
\doival{10.1017/xxxxx}

\aopheadtitle{Proceedings IAU Symposium}
\editors{Hyung Mok Lee, Rainer Spurzem and Jongsuk Hong, eds.}

\title{GW emission and relativistic dynamical friction in intermediate mass ratio inspirals}

\author{P. Di Cintio$^{1,2,3}$, G. Bertone$^4$, C. Chiari$^{5,6}$, T. K. Karydas$^4$,\\
B. J. Kavanagh$^7$, M. Pasquato$^{8,9}$, A. A. Trani$^{10,11,12}$}
\affiliation{$^1$CNR-Istituto dei Sistemi Complessi, Via Madonna del piano 10, I-50019 Sesto Fiorentino, Italy}
\affiliation{$^2$INFN-Firenze, via G.\ Sansone 1, I-50019 Sesto Fiorentino, Italy}
\affiliation{$^3$INAF - Osservatorio Astrofisico di Arcetri, largo E.\ Fermi 5, I-50125, Firenze, Italy}
\affiliation{$^4$Gravitation Astroparticle Physics Amsterdam (GRAPPA),\\ Institute for Theoretical Physics Amsterdam and Delta Institute for Theoretical Physics,\\ University of Amsterdam, Science Park 904, 1098 XH Amsterdam, The Netherlands}
\affiliation{$^5$Dipartimento di Scienze Fisiche, Informatiche e Matematiche, Universit\'a di Modena e Reggio Emilia, Via Campi 213/A, I-41125 Modena, Italy}
\affiliation{$^6$CNR-NANO, via Campi 213/A I-41125, Modena, Italy}
\affiliation{$^7$Instituto de F\'isica de Cantabria (IFCA, UC-CSIC), Av.~de Los Castros s/n, 39005 Santander, Spain} 
\affiliation{$^8$INAF-IASF Via Alfonso Corti 12 I-20133 Milano, Italy}
\affiliation{$^9$INAF - Osservatorio Astronomico di Padova, Vicolo dell'Osservatorio 5, I-35122 Padova, Italy}
\affiliation{$^{10}$Niels Bohr International Academy, Niels Bohr Institute, Blegdamsvej 17, 2100 Copenhagen, Denmark}
\affiliation{$^{11}$INFN-Trieste, I-34127, Trieste, Italy}
\affiliation{$^{12}$Departamento de Astronom\'ia, Facultad Ciencias F\'isicas y Matem\'aticas, Universidad de Concepci\'on, Avenida Esteban Iturra, Casilla 160-C, Concepci\'on, 4030000, Chile}
%\pubyear{2025}
%\volume{398}  
%\setcounter{page}{1}
%\jname{Compact Stars and Binaries in Dense Star Clusters}
%\editors{Hyung Mok Lee, Rainer Spurzem and Jongsuk Hong, eds.}
%%%%%%%%%%%%%%%%%%%%%%%%%%%%%%%%%%%%%%%%%%%%%%%%%%%%%%%%%%%%%%%%%%%%%%%%
\begin{abstract}
We present a set of preliminary simulations of intermediate mass ratio inspirals (IMRIs) inside dark matter (DM) spikes accounting for post-Newtonian corrections the interaction between the two black holes up to the order 2.5 in $c^2$, as well as relativistic corrections to the dynamical friction (DF) force exerted by the DM distribution. We find that, incorporating relativity reduces of a factor $1/2$ the inspiral time, for equivalent initial orbital parameters, with respect to the purely classical estimates. Vice versa, neglecting the DF of the spike systematically yields longer inspiral times.  
\end{abstract}
%\firstsection 
\maketitle
%%%%%%%%%%%%%%%%%%
\section{Introduction}
Massive black holes (BHs) are likely to be sitting at the centre of dark matter (DM) structures such as spikes or mounds that may or may not evaporate across cosmic time according to the specific nature of DM itself or the black hole mass.\\ 
\indent It has been speculated that future gravitational waves (GW) observatories can probe the DM distribution around intermediate or supermassive BHs with masses $M_{\rm BH}$ ranging from $10^2$ to $10^{9}{\rm M}_\odot$, and possibly obtain constraints on the nature of DM, by detecting the dephasing in the waveform of binary BH mergers induced by said DM overdensities (see \citealt{2020PhRvD.102h3006K}). The accurate modelling of the dynamical friction (\citealt{Chand_DF}, DF), induced by the DM on the secondary compact object of mass $m_{\rm BH}\approx 1{\rm M}_\odot$ in intermediate (IMRI) and extreme mass ratio inspirals (EMRI) is therefore of pivotal importance.\\
\indent In \cite{2025PhRvD.111f3070K} and \cite{2025PhRvD.111f3071K} we introduced a simple semi-analytical scheme to account for the effect of DF and the back reaction of the inspiralling compact object on the DM distribution and demonstrate that including stirring (i.e. cusp density fluctuations induced by the time-dependent potential of the binary BH) tends to slow the rate of dark matter depletion and therefore enhances the impact of the drag exerted by DM on the dynamics of the binary. Additionally, we validated our models using the numerical code {\sc NbodyIMRI}, a new publicly available code designed for simulating binary systems within DM spikes (\citealt{2024ascl.soft04020K}) that models the DM distribution as a set of $N$ non-interacting simulation particles with tunable mass and force acting on $m_{\rm BH}$.\\  
\indent So far, our treatment was purely classical, while it is of course natural that, in particular in the last phases of the DF driven inspirals, relativistic effects do play a non-negligible role. Here we introduce another simple numerical code incorporating corrections to both the DF induced by the DM spike and the force field of the central massive BH.  
\section{Models}
Following the approach of \cite{PhysRevD.106.064003} and \cite{2025arXiv250602173T}, we integrate the orbit of the smaller companion $m_{\rm BH}$ in the central field of the massive BH $M_{\rm BH}$, accounting for the effect of DF as parametrized by the friction coefficient $\eta$ given by
\begin{equation}\label{eq1}
    \frac{{\rm d}^2\mathbf{r}}{{\rm d}t^2}=\mathbf{F}_{\rm tot}-\eta\mathbf{v},
\end{equation}
using the \cite{mik06} modified midpoint leapfrog scheme with adaptive timestep. In Eq. (\ref{eq1}) $\mathbf{F}_{\rm tot}=\mathbf{F}_{\rm BH}-\nabla\Phi_{\rm DM}$ is the total force per unit mass. The contribution of the central BH is
\begin{equation}
  \mathbf{F}_{\rm BH}=-\frac{GM_{\rm BH}}{r^2}\bigg[(1+\mathcal{A})\frac{\mathbf{r}}{r}+\mathcal{B}\mathbf{v}\bigg],
\end{equation}
with the coefficients of the post-Newtonian (PN) expansion in powers of the speed of light $c$, $\mathcal{A}$ and $\mathcal{B}$, summed up to the order 2.5 (for the explicit expression see e.g. \citealt{2004PhRvD..69j4021M}), corresponding to the lowest order accounting for energy loss in GW emission. The DM cusp generates the potential $\Phi_{\rm DF}$ via the usual (classical) Poisson equation associated to the density 
\begin{equation}\label{rhodm}
\rho_\mathrm{DM}(r) = \rho_6 \left(\frac{r}{r_6}\right)^{-\beta} \left( 1 + \frac{r}{r_t}\right)^{-\alpha},
\end{equation}
where $r_6 = 10^{-6}\,\mathrm{pc}$ is a reference radius; $\rho_6$ specifies the density normalization (at said radius); $\alpha$ and $r_t$ are the slope and the cutoff radius of the truncation, and $\beta$ is the inner density slope.\\
\indent The DM density also produces DF on the companion $m_{\rm BH}$, here we use the approximate form of the relativistic coefficient evaluated in \cite{2023A&A...677A.140C}
\begin{equation}
    \eta\approx 4\pi G^2m_{\rm BH}\rho_\mathrm{DM}(r)\frac{\log\Lambda}{\sigma^3\gamma^3},
\end{equation}
where $\gamma=(1+v^2/c^2)^{-1/2}$ and we have assumed that the velocity dispersion $\sigma$ of the DM particles in the cusp locally equals the the circular velocity in the (classical) Keplerian potential of $M_{\rm BH}$. The argument of the Coulomb logarithm $\Lambda=b_{\rm max}/b_{\rm min}$ is constrained as in \cite{2025PhRvD.111f3071K} with $b_{\max}\approx r/3$ and $b_{\rm min}\approx Gm_{\rm BH}/v^2$.     
\begin{figure}
  \includegraphics[width=0.7\textwidth]{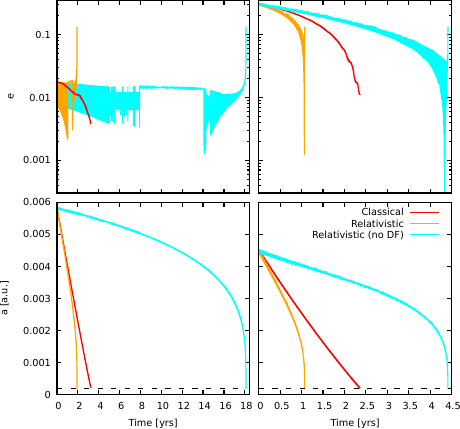}
\caption{Evolution of the orbital eccentricity $e$ (top panels) and the semimajor axis $a$ (bottom panels) in the last phase of a IMRI for a 20$M_\odot$ black hole orbiting an IMBH of $10^3M_\odot$ on a nearly circular orbit (left panels) and a moderately eccentric orbit (right panels).}
\label{eanda}       
\end{figure}
\section{Preliminary results and implications}
\begin{figure}
  \includegraphics[width=0.8\textwidth]{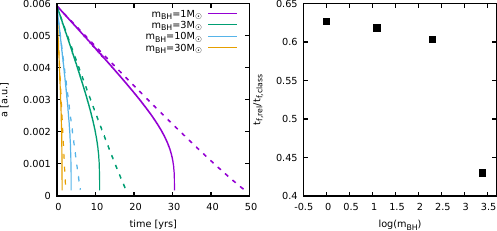}
\caption{Evolution of the semimajor axis $a$ in the last phase of an IMRI for different black hole masses ($m_{BH}=1,$ 3, 10 and $30{\rm M}_\odot$) orbiting an IMBH of $10^3{\rm M}_\odot$ on a (initially) nearly circular orbit (left panel). The solid lines refer to the relativistic simulations with up to 2.5PN terms while the dashed lines mark the parent purely classical integration. Ratio of the inspiral time down to $10r_S$ in relativistic and classical simulations as function of $m_{BH}$ (right panel).}
\label{rvsc}       
\end{figure}
In the same spirit of our previous work with $N-$body simulations, we integrated the orbit of the companion around an intermediate mass BH with $M_{\rm BH}=10^3{\rm M}_\odot$, with $m_{\rm BH}$ ranging from one to $30{\rm M}_\odot$. We fixed the cusp parameters in Eq. (\ref{rhodm}) to $\alpha=3$; $\beta=2.33$ and $\rho_6=10^{16}{\rm M_\odot {\rm pc}^{-3}}$. For this specific choice of cusp parameters, the associated gravitational potential is less than one part in $10^2$ at a separation of $10^{-3}$ astronomical units, so that its contribution to $\mathbf{F}_{tot}$ could be in principle neglected. We considered different values of the orbital eccentricity $e$, from a moderately eccentric orbits with $e=0.5$ to circular orbits. All integrations of Eq. (\ref{eq1}) where extended up to the time at which the binary separation reaches ten times the Schwartzschild radius of the central BH $r_S=2GM_{\rm BH}/c^2$.\\
\indent Figure \ref{eanda} shows the evolution of $e$ and the semimajor axis $a$ for two sample initial conditions and three different simulation set-ups. Namely the purely classical case (i.e. $\mathcal{A}=\mathcal{B}=0$; $\gamma=1$), in a way corresponding to the simulations in {\sc NbodyIMRI}; the full relativistic case, and an additional relativistic case where the contribution of the DM cusp is neglected for both DF and potential.\\
\indent We observe that for non vanishing $e$ both DF and GW emission-induced dissipation tend to circularize the orbit, however at about $r\approx20r_S$ the PN corrections to the gravitational acceleration result in a drastic increase of the eccentricity. This latter behavior is expected because of the Newtonian definition of eccentricity that we employ \citep[see][and reference therein]{2019CQGra..36s5013W}. If on one hand in classical DF-driven inspirals the decay of the semimajor axis $a$ is always nearly linear, once the PN terms are dominant the decay rate increases with a steep power-law resulting in a reduction of the infall time $t_f$ of a factor of about $1/2$, as shown in Figure \ref{rvsc}  for a circular BH of mass 1, 3, 10 and 30 ${\rm M}_\odot$ on an initially circular orbit of semimajor axis $6\times 10^{-3}$ astronomical units. Larger values of the companion's mass result in stronger relativistic corrections (since $m_{\rm BH}$ enters explicitly in the definitions of $\mathcal{A}$ and $\mathcal{B}$) and hence $t_f$ is further reduced with respect to the classical case for analogous initial orbital parameters.\\
\indent As a general trend, we confirm the result of our $N-$body simulations (and those of similar works such as \citealt{2024MNRAS.533.2335M}) that the presence of a DM spike surrounding the primary component of the binary significantly enhances the inspiral (cfr cyan and orange curves in Fig.~\ref{eanda}), even in the cases when its gravitational field is negligible. We stress the fact that our simple model could be also applied to a broad range of mass ratios $m_{\rm BH}/M_{\rm BH}$, in principle allowing to simulate EMRI in galactic cores. This established, the observability of dynamical friction effects in the gravitational wave band remains to be determined, which will be addressed in our follow-up work. Moreover, as the BH mass ratios explored here are likely to be associated to a non-negligible DM profile evolution (see e.g \citealt{2022PhRvD.105d3009C}), an extension of the numerical method that incorporates this feature is currently in the works. In particular, this approach would become relevant to probe alternative models of DM, such as for example  the superfluid DM, where DF departs significantly from the Chandrasekhar formulation (\citealt{2024JCAP...06..024B}).

\bibliographystyle{iaulike}
\bibliography{main} 
\end{document}